# Mitochondria in higher plants possess $H_2$-evolving activity which is closely related to complex Ⅰ

Xin Zhang[1,2]†, Zhao Zhang[1,2]†, Yanan Wei[1], Muhan Li[1], Pengxiang Zhao[1,2], Yao Mawulikplimi Adzavon[1,2], Mengyu Liu[1,2], Xiaokang Zhang[1,2], Fei Xie[1,2], Andong Wang[3], Jihong Sun[3], Yunlong Shao[3], Xiayan Wang[3], Xuejun Sun[4] & Xuemei Ma[1,2]*.

[1]College of Life Science and Bioengineering, Beijing University of Technology, Beijing 100124, P. R. China;

[2]Beijing molecular hydrogen research center, Beijing 100124, P. R. China;

[3]College of Environmental and Energy Engineering, Beijing University of Technology, Beijing 100124, P. R. China;

[4]Department of Navy Aeromedicine, The Second Military Medical University, Shanghai 200433, P. R. China.

*Correspondence to: xmma@bjut.edu.cn

† These authors contributed equally to this work.

**Abstract:**
Hydrogenases occupy a central place in the energy metabolism of anaerobic bacteria. Although the structure of mitochondrial complex Ⅰ is similar to that of hydrogenase, whether it has hydrogen metabolic activity remain unclear. Here, we show that a "$H_2$-evolving activity" exists in higher plants mitochondria and is closely related to complex Ⅰ, especially around ubiquinone binding site. The $H_2$ production could be inhibited by rotenone and ubiquinone. Hypoxia could simultaneously promote $H_2$ evolution and succinate accumulation. Redox properties of quinone pool, adjusted by NADH/succinate according to oxygen concentration, acts as a valve to control the flow of protons/electrons and the production of $H_2$. The coupling of $H_2$-evolving activity of mitochondrial complex I with metabolic regulation reveals a more effective redox homeostasis regulation mechanism. Considering the ubiquity of mitochondria in eukaryotes, $H_2$ metabolism might be the innate function of higher organisms. This may serve to explain, at least in part, the broad physiological effects of $H_2$.

## Introduction

There have been several reports on hydrogen ($H_2$) metabolism in higher plants. In 1961, Sanadze (Sanadze, 1961) firstly provided evidence that the leaves of higher plants could both release and absorb $H_2$. Thereafter, many kinds of higher plants, e.g. *barley* (Torres et al., 1986), *Arabidopsis* (Xie et al., 2014), *rice* (Xu et al., 2013) *and Medicago sativa* (Jin et al., 2016), have been reported to produce $H_2$ during seed's germination or from tissue lysis when plants faced abiotic stresses. The photosystem (PS II or PSI) in chloroplasts had been shown to have $H_2$ evolution activity (Mal'tsev et al., 1988). However, during seed's germination, there was no chloroplast formation, suggesting that there should be other components involved in $H_2$ production.

Hydrogenases, mainly found in bacteria, archaea and algae, are metalloenzymes that catalyze the reversible reaction of dihydrogen into protons and electrons: $H_2 \leftrightharpoons H^+ + H^- \leftrightharpoons 2H^+ + 2e^-$ (Lubitz et al., 2014). The atmosphere of the early Earth was hydrogen-rich, it is reasonable to think that hydrogenases were probably "invented" during the earliest life on our planet (Cammack, 1999; Vignais & Billoud, 2007). Eukaryotes are suggested to have arisen through symbiotic association of an anaerobic, strictly hydrogen-dependent, strictly autotrophic archaebacterium (the host) with a eubacterium (the symbiont) that was able to respire, but generated molecular $H_2$ as a waste product of anaerobic heterotrophic metabolism. $H_2$ is the bond that forges eukaryotes out of prokaryotes(Martin & Müller, 1998). If this is the truth, the eukaryotes symbiont's $H_2$-metabolism function might be not lost and be remained in mitochondria so as to be readily recruitable.

Complex I (NADH:ubiquinone oxidoreductase, EC 1.6.99.3), found in the respiratory chain of mitochondria or bacteria of all kingdoms of life, catalyzes the oxidation of NADH and the reduction of ubiquinone, coupled with the translocation of protons across the membrane(Marreiros et al., 2013). There have been many reports on the homology and evolutionary relationship of mitochondrial complex I and hydrogenase (Efremov & Sazanov, 2012). The core of the enzyme complex consist of 14 "core" subunits (7 hydrophilic and 7 hydrophobic) conserved from the prokaryote to eukaryote mitochondria. The energy-converting hydrogenase (Ech), which is present in many anaerobic or facultative anaerobic microorganisms, showed high sequence similarity to complex I core subunits (Efremov & Sazanov, 2012). Moreover, complex I also relates to a class of membrane-bound NiFe hydrogenases coupling substrate oxidation and reduction of hydrogen to active proton transport (Hedderich, 2004). The quinone binding site in complex I appears to correspond to the NiFe active site in hydrogenase (Brandt, 2006; Tocilescu et al., 2010). But until now, there is no experimental evidence that mitochondria of eukaryotes possess $H_2$ metabolism activities.

## Results

### $H_2$-evolving activity existed in mitochondria of higher plants

We used headspace gas chromatography (GC) method to detect the $H_2$ content (Figure S1). The anaerobic environment was achieved by aeration of nitrogen gas ($N_2$) in the sealed headspace bottle, which could exclude the gas exchange with the air and easily monitor the production of $H_2$ and consumption of oxygen ($O_2$). Shaking or not of the sealed bottles had no effect on the detection of $H_2$ (Figure S2A).

When sealed in headspace bottle, $H_2$ production during *Vigna radiata* seedlings growth in both normoxic and hypoxic treatments was found (Figure 1A). For more than 24 hours, the $O_2$ concentration was around 7% in normoxic treatment (lower than 21% $O_2$ in the air) and more than 10 times lower in hypoxic treatment (Figure 1B), indicating that low $O_2$ concentration may active the hydrogenase-like activity which was similar with previous study(Torres et al., 1986). Roots, hypocotyls, cotyledons and leaves all had the ability to release $H_2$ (Figure 1C).

To explore the sources of $H_2$-evolving hydrogenase activity other than chloroplasts, *V. radiata* seedlings hypocotyls were chosen as starting materials. The mitochondria or plasma membrane-rich crude lysis of 72h-seedlings hypocotyls separated by differential centrifugation have high $H_2$ production activity (Figure S2B). To test whether mitochondria really own hydrogenase-like activity, crude mitochondria of *V. radiata* hypocotyls were extracted and it was found that buffer composition especially metal ion chelators had great impact on $H_2$ production (Figure S2C). When dissolved in washing medium (0.3 M mannitol, 0.1% (w/v) BSA, 10 mM MOPS, pH7.2), the $H_2$ evolution activity of the crude mitochondrial was significantly inhibited by EDTA (Figure 2D) which suggested an metallo-hydrogenase might exit. The optimal conditions for $H_2$ evolution were pH 6.0 and 35 °C (Figure 1E and 1F).

Most of the hydrogenases found in microorganism and algae were $O_2$ sensitive (Horch et al., 2012). To study the effects of $O_2$ on $H_2$ production, mitochondria solution was sealed in the headspace bottle. As shown in Figure 1H, mitochondria had the ability to produce $H_2$ whether it is normoxia (21% $O_2$) or hypoxia (~5% $O_2$) under the seedlings growing temperature (i.e. the *V. radiate* germination and growth temperature at 25±1°C), and the $H_2$ evolution both increased with the prolongation of incubation time. Promoted $H_2$ evolution by hypoxia was observed (Figure 1H and 1I), mitochondria released about 39 nmol $H_2$/ mg protein at 6 h, and the cumulative amounts of $H_2$ increased to 556 nmol $H_2$/ mg protein at 8 h. Under normal $O_2$ condition, mitochondria began to produce 29 nmol $H_2$/ mg protein at 8 h, with a relative delay of 2 hours (Figure 1I). The evolution of $H_2$ could be significantly promoted and the time was shortened to 3 hours at the optimal temperature (35°C) under hypoxia and reached a maximum of 6060 nmol / mg protein at 7 h (Figure 1J). The mitochondria of other higher plants, such as *Arabidopsis thaliana* and *Glycine max,* also had the ability to produce $H_2$ under hypoxia, with a cumulative amount of 34 nmol $H_2$/mg protein and 699 nmol $H_2$/mg protein at 6h, respectively (Figure 1G). These results prompted an evidence that an $O_2$ sensitive $H_2$-evolving hydrogenase activity existed in mitochondria of higher plants.

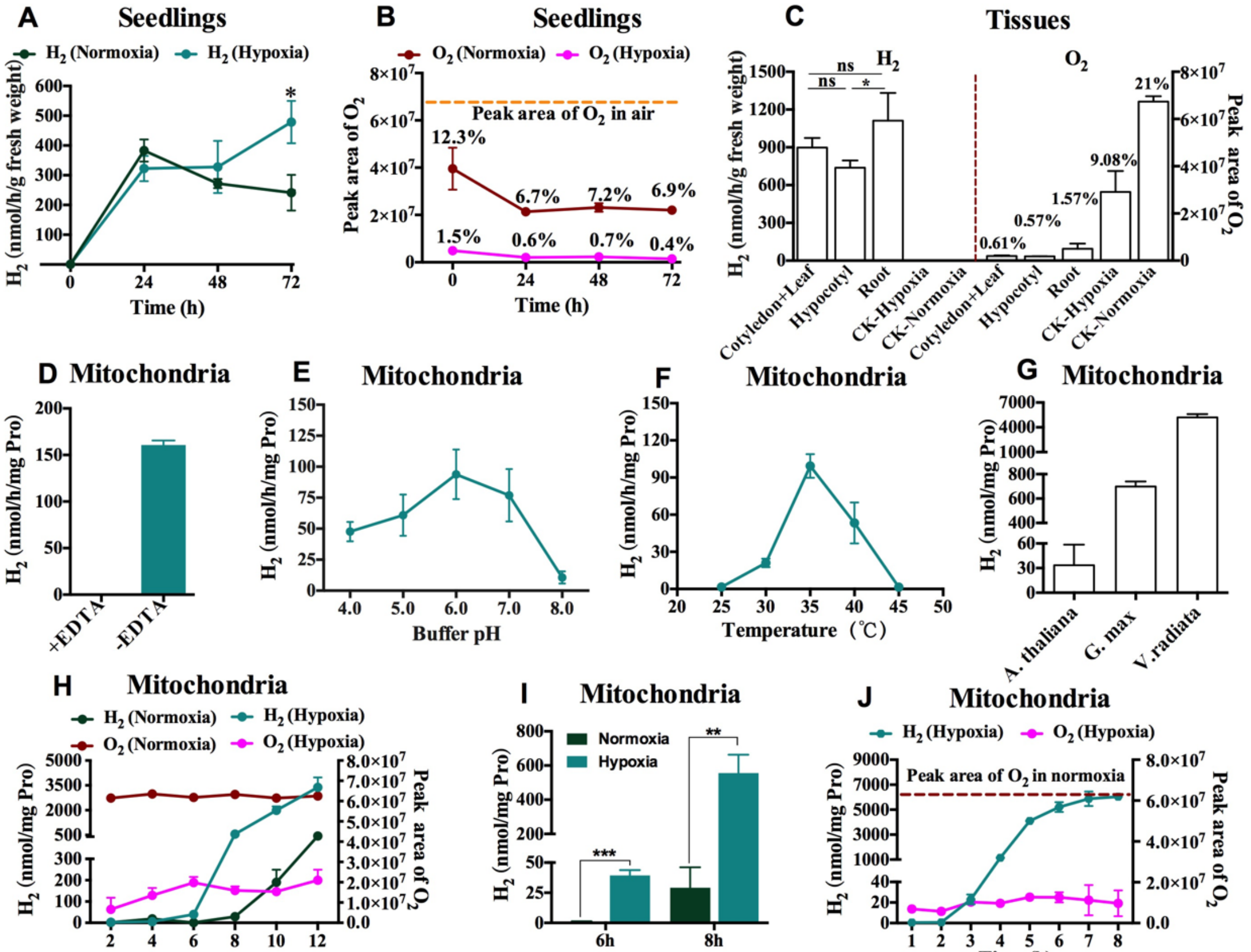

**Figure 1** $H_2$ evolution of seedlings, tissues or crude mitochondria of high plants. A-B, Time course of $H_2$ evolution (A) and $O_2$ consumption (B) by *V.radiata* seedlings (n=5). C, $H_2$ evolution and $O_2$ consumption by *V.radiata* tissues. CK, the blank control. The numbers on the broken lines or columns indicated the percentage of $O_2$ (n=4). D, The influence of EDTA on $H_2$ evolution activities of mitochondria (n=3). E-F, The optimal pH (E) (n=3) or temperature (F) (n=2) of $H_2$ evolution activities in crude mitochondria. G, The $H_2$ evolution activity of *A. thaliana, G. max* and *V. radiata* mitochondria for 6 hours at 35℃ (n=2 for *G.*

*max* and n=3 for others). H, The $H_2$ evolution activity of *V. radiata* mitochondria after different times of hypoxic or normoxic treatment at 25℃ (n=3). I, The $H_2$ cumulative amount of *V.radiata* mitochondria for 6 hours or 8 hours at 25℃ (n=3). J, The $H_2$ evolution activity of *V. radiate* mitochondria after different times of hypoxic treatment at 35℃ (n=3). Data shown are the means ± SD. ns, no significant difference ($p > 0.05$), * $P < 0.05$, **$P < 0.01$, ***$P < 0.001$.

## Metabolite changes during $H_2$ evolution

But why is $H_2$ generation delayed for several hours? What happened during the process? To open the black box of $H_2$ metabolism, the metabolomics changes in mitochondria during $H_2$ evolution at 25℃ by crude mitochondria of *V. radiata* hypocotyls were studied. The tricarboxylic acid (TCA) cycle metabolic intermediates changed a lot, especially succinate elevated by as much as 100-fold over hypoxic periods of 10 hours and 2-fold higher than that under aerobic condition (Figure S3B) which is consistent with $H_2$ evolution (Figure S3A).

TCA cycle is a nearly universal central catabolic pathway in eukaryote mitochondria with most of the energy of oxidation temporarily held in the electron carriers $FADH_2$ and NADH (Figure 2A). During aerobic metabolism, these electrons are transferred into electron transfer chain (ETC) to reduce $O_2$ to $H_2O$ and the energy of electron flow is trapped as ATP (Figure 4A). Chouchani and colleagues have demonstrated that once subjected to hypoxia, a truncated electron transfer between complex I and II (Succinate dehydrogenase, SDH) occurred, and the accumulation of TCA intermediate succinate via fumarate production and reversal of complex II is an universal metabolic signature of ischaemia (hypoxia) in vivo(Chouchani et al., 2014). The question is whether $H_2$ production promoted by hypoxia also related to succinate accumulation?

To further conform the relationship of mitochondrial metabolism and $H_2$ evolution, different TCA metabolic intermediates were added into the reaction system. Pyruvate and malate, fuel of the complex I, promoted the production of $H_2$ by 10 times which could be greatly inhibited by rotenone (Figure 2B). Succinate and fumarate, substrate of the complex II, significantly promoted the production of $H_2$ which could be greatly inhibited by malonic acid (Figure 2C). Mitochondrial ETC complex I/II are the two primary link points to TCA cycle, which are involved in the oxidation of NADH and succinate respectively. (Figure 2A). These results indicated that complex I and II were closely related to $H_2$ evolution.

It's possible that the accumulation of succinate promotes the reverse transfer of electrons from complex II to I and the generation of $H_2$. To test this hypothesis, Thermo Orbitrap MS was used to investigate the time changes of TCA metabolites during $H_2$ evolution at 35℃ (Figure S4). The production of $H_2$ was evidently accompanied by succinate accumulation and the process could be divided into three stages (Figure 2D-I). In the early stage (0-1h, stage I), there was no $H_2$ produced. Succinate, fumarate and malate decreased while α-ketoglutarate increased, indicating that the metabolic rate of TCA decreased because of the low $O_2$ concentration (Figure 2F-2I). In stage II (1-3h), malate, fumarate and succinate increased gradually in time sequence (Figure 2G-2I), which suggest a reverse pathway from α- ketoglutarate via malate and fumarate to succinate(Chouchani et al., 2014). Stage III(4-7h) represented the period of high-speed $H_2$ evolution, an obvious accumulation of α- ketoglutarate and succinate occurred, while the malate, fumarate remained a relatively low level, indicating that the normal TCA cycle from α- ketoglutarate to succinate is dominant again (Figure 2F, 2G). Succinate was consistently elevated by as much as 55-fold in Stage III than 0 hour (Figure 2G), indicating that it was a primary driver of the mitochondrial $H_2$ production. So we confirmed that the $H_2$ evolution was relative to the succinate accumulation caused by the reverse electron transfer. In addition to succinate, the increased NADH content also was found with the mitochondria incubation under hypoxic conditions in stage III and reached to 1.95-fold higher than 0 hours (Figure 2E).

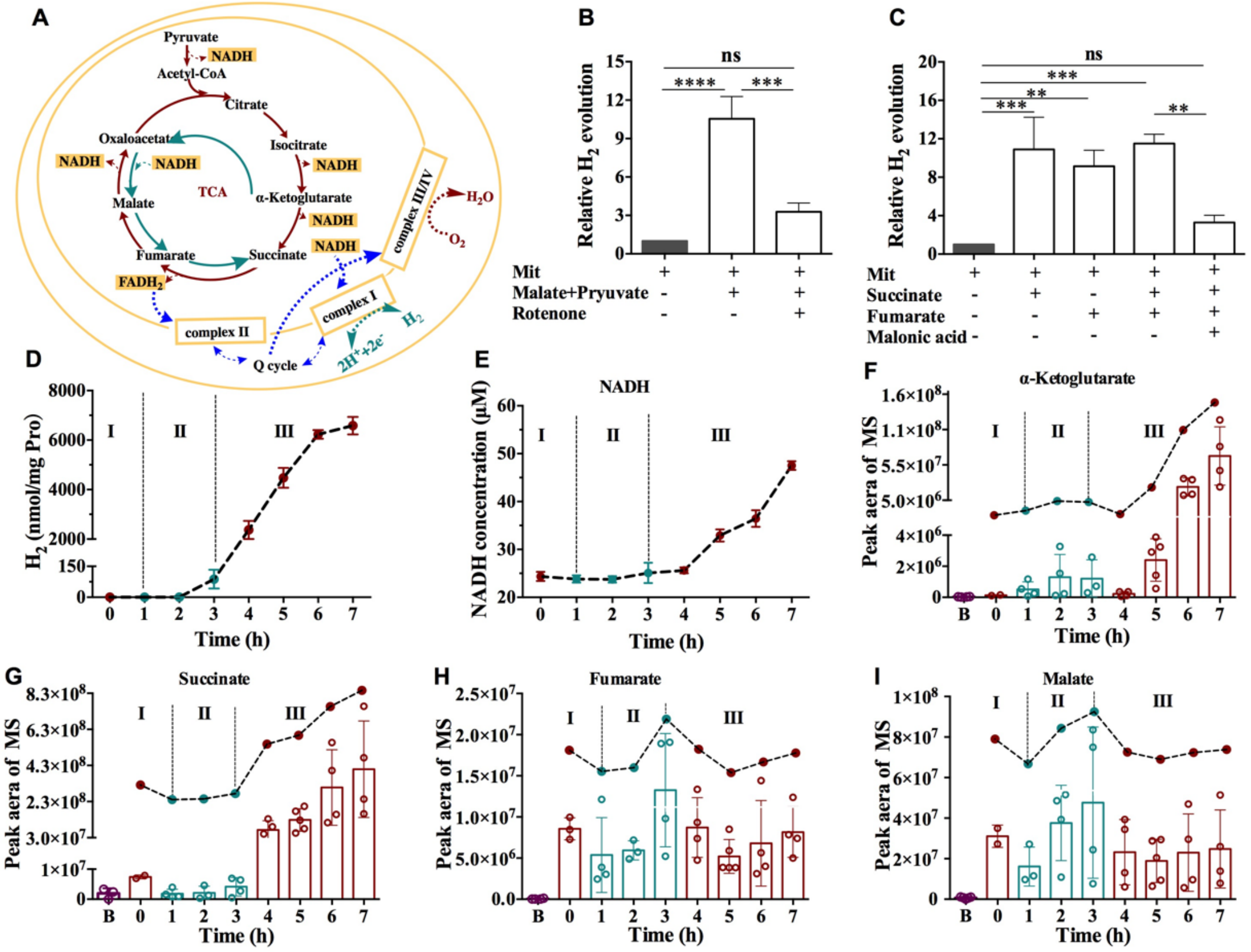

Figure 2 Metabolite changes during $H_2$ evolution under hypoxia. A, The schematic diagram of TCA cycle during $H_2$ evolution. B-C, The relatioship between $H_2$ metabolism and complex I (B), II (C) (mean±SD, n=3). D, The time-course of $H_2$ evolution in hypoxia at 35℃ (mean±SD, n=3). E, The change of NADH content during $H_2$ evolutionTT (mean±SD, n=3 or 4). F-I, Analysis of α-ketoglutarate (E), succinate (F), fumarate (G) and malate (H) changes during $H_2$ evolution. Each hollow circle in the column represents one sample (mean±SD, n=2-5). Line chart represents the trend of metabolites. ns, no significant difference ($p > 0.05$), **$P < 0.01$, ***$P < 0.001$, ****$P < 0.0001$.

## $H_2$-evolving activity closely related to mitochondrial complex I

To further explore the exact site of the $H_2$-evolving hydrogenase-like activity, ETC inhibitors (Figure 3A) were added into the reaction system. The mitochondrial complex I inhibitor rotenone, complex IV (cytochrome c oxidase) inhibitor $NaN_3$ almost completely inhibited the mitochondrial $H_2$ production, respectively (Figure 3B). Only 18% or 38% of $H_2$ production were blocked by complex II inhibitor malonic acid or complex III (ubiquinone: cytochrome c oxidoreductase) inhibitor antimycin A, respectively (Figure 3B). Above results suggested that the electron transfer pathway I → III was more closely correlated with $H_2$ metabolism compared with II → III. Rotenone, a complex I inhibitor, which is known to inhibit the downstream eT from N2 (the terminal FeS cluster) cluster to ubiquinone(Haapanen & Sharma, 2018), almost completely blocked the $H_2$ evolution even when pyruvate and malate were added (Figure 3B, 2B), suggesting the close relationship of $H_2$ evolution with complex I. Rather, the addition of oligomycin, which binds to a 23 kd polypeptide in the F0 subunit of the F0/F1 ATPase (ATP synthetase, complex V) thereby preventing protons from passing back into the mitochondria matrix, promoted $H_2$ production (Figure 3B). The uncoupler FCCP (carbonyl cyanide- 4-(trifluoromethoxy) phenylhydrazone), which destroyed the protons electrochemical gradient of the inner membrane almost completely inhibited the mitochondrial $H_2$ production (Figure 3B). The pH of the reaction system gradually decreases with time (Figure 3C), indicating that protons accumulated on the outer side of the inner membrane was important to the formation of $H_2$.

But how do complex I, II and protons coordinately produce $H_2$ remain unclear. Ubiquinone can exist in three different redox states: fully oxidized (UQ), partially reduced (ubisemiquinone, $UQ^{\cdot-}$), and fully reduced ($UQH_2$). In the inner mitochondrial membrane, UQ dynamically shuttles between complexes I, II and III, transferring electrons coupled with the uptake and release of protons(Wang & Hekimi, 2016). We found that the addition of 60 μM UQ blocked $H_2$ production completely (Figure 3D). When mixed TCA intermediates (5mM pyruvate+5mM malate+5mM α-ketoglutarate+10mM succinate+5mM fumarate) were added to provide sufficient substrate for ETC, the production of $H_2$ increased by 20-fold which could be significantly inhibited by UQ. Rotenone and malonic acid could further inhibit the $H_2$ production, which was only 2.6% and 68% of the UQ plus mixed acid group, respectively (Figure 3D). The above results is consistent with previous speculation that the UQ binding site in complex I, appears to correspond to the NiFe active site in hydrogenase (Figure S5) (Moparthi & Hagerhall, 2011) and the $H_2$-evolving hydrogenase-like activity should be located in mitochondrial complex I, especially around UQ binding site. The adequate sources of electron and proton, the coordination of complexes I and II metabolism to inhibit the oxidation of $UQH_2$ to UQ, serves as important preconditions in $H_2$ production.

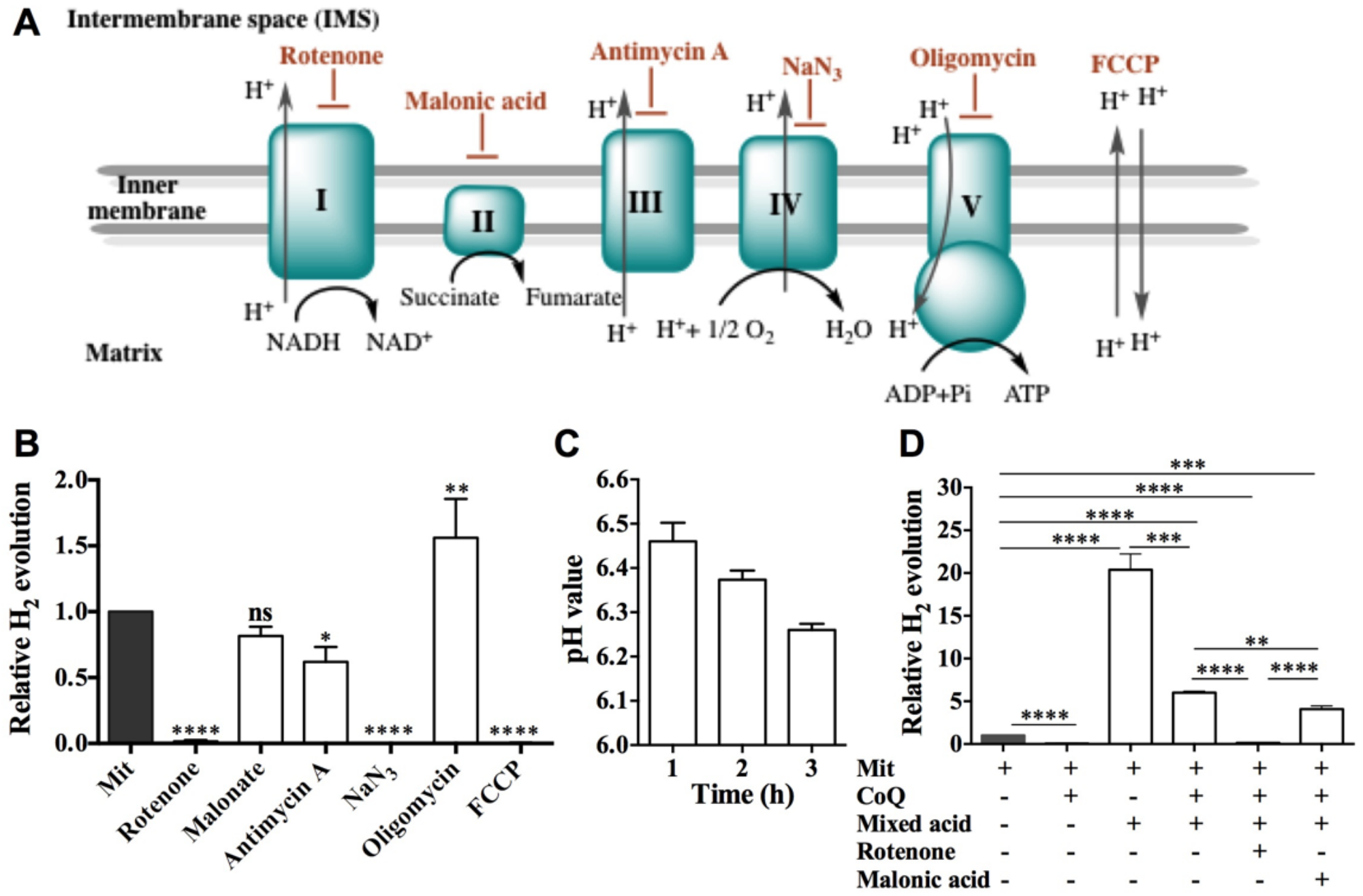

**Figure 3 $H_2$ evolution by mitochondria was closed related to ETC complex I.** A, The schematic diagram of the target action of ETC inhibitors. B, The $H_2$ evolution was influenced by different inhibitors (mean±SD, n=3). C, pH changes during $H_2$ evolution (mean±SD, n=2). D, The relatioship between $H_2$ metabolism and UQ (mean±SD, n=3). Mixed acid: 5mM pyruvate+5mM malate+5mM α-ketoglutarate+10mM succinate+5mM fumarate. ns, no-significant, *$P < 0.05$, **$P < 0.01$, ***$P < 0.001$, ****$P < 0.0001$.

## $H_2$ evolution mechanism in mitochondria

$H_2$ might be formed in a hydrogen-rich, reduction environment, the following conditions could inhibit $H_2$ production: continuous destruction of reducibility e.g. by $O_2$, protons and electrons can not be combined, or two hydrogen atoms can not be close to each other. The standard reduction potentials of UQ/$UQH_2$ is 0.045mV, while $H^+/H_2$ is -0.414mV. Electrons tend to flow to the higher reduction potential, and the strength of this tendency is not only proportional to the difference between the two reduction potentials ($\Delta E$) but also a function of the concentrations of oxidized and reduced species. During aerobic metabolism, electrons from NADH/succinate are transferred into ETC (Figure 4A), protons are weak in competing with UQ to accept electrons. But during anaerobic metabolism, the oxidation of $UQH_2$ to UQ is inhibited, the accumulated protons have the chance to meet electrons and produce $H_2$.

In recent years, increasing evidence identified that there were two UQ binding sites in complex I(Haapanen et al., 2019; Wang & Hekimi, 2016). The $UQ^-$ may act as a strong reductant to drive the reduction of UQ at the second UQ binding site ($Q_{2nd}$) to form $UQH_2$, then $UQH_2$ left the chamber to continue the subsequent redox process(Haapanen & Sharma, 2018). Rotenone blocks the entrance of UQ-chamber, protons and UQ can't access and the $H_2$ production were completely inhibited (Figure 3B, Figure 3D). UQ could also inhibit the $H_2$ production (Figure 3D), suggesting that around the $Q_{2nd}$ is likely where rotenone, UQ and protons competitively react. We present a perspective that around the mitochondrial complex I N2 cluster especially $Q_{2nd}$ is the central of the reaction "arena", where protons/$H_2$, quinone (Q) species (UQ, $UQ^-$, $UQH_2$) are likely to compete to donate or accept protons/electrons which closely related to $H_2$ metabolism (Figure 4F). Redox properties of Q pool, adjusted by NADH/succinate according to $O_2$ concentration, acts as a valve to control the flow of protons/electrons and the production of $H_2$. We propose the following 3-stage $H_2$ evolving process (Figure 4C-4E):

1. Under normoxia or in the early stage of hypoxia (stage I), the TCA cycle metabolism is in the normal direction and α- ketoglutarate is oxidized to succinate with the metabolic activity decreased in the early stage of hypoxia. Electrons, comes from NADH, are transferred along the chain of FeS clusters to N2 cluster, and further transferred to $Q_{1st}$ (the first UQ-binding site where produce $UQ^-$) and $Q_{2nd}$ (the second UQ-binding site where produces $UQH_2$) to form $UQH_2$ (Figure 4C), no $H_2$ produced.

2. During hypoxia (stage II), fumarate production increases, a short electron transport pathway, via complex I→$UQH_2$→fumarate, promote the reversal of complex II and accumulation of succinate (Figure 4B and 4D). At the end point of stage II, the accumulated succinate feedback inhibits complex II, $UQH_2$ accumulated and the oxidized UQ is lacking, the accumulated protons have more competitive advantages than UQ at $Q_{2nd}$ to accept electrons transferred from N2 cluster and began to generate $H_2$ (Figure 4D).

3. During prolonged hypoxia (stage III), the normal TCA cycle from α ketoglutarate to succinate is dominant again and succinate is obviously elevated because of the inhibition of complex II, mass production of $H_2$ occurred. With the increased NADH/$NAD^+$ ratio and the accumulated succinate is oxidized to maintain the Q pool reduced, thereby sustaining a large electromotive force by electron transport from NADH to N2 clusters but stopped for lacking of UQ. Protons instead accept the electrons to produce $H_2$, and the $H_2$ evolution speed reach maximum (Figure 4E).

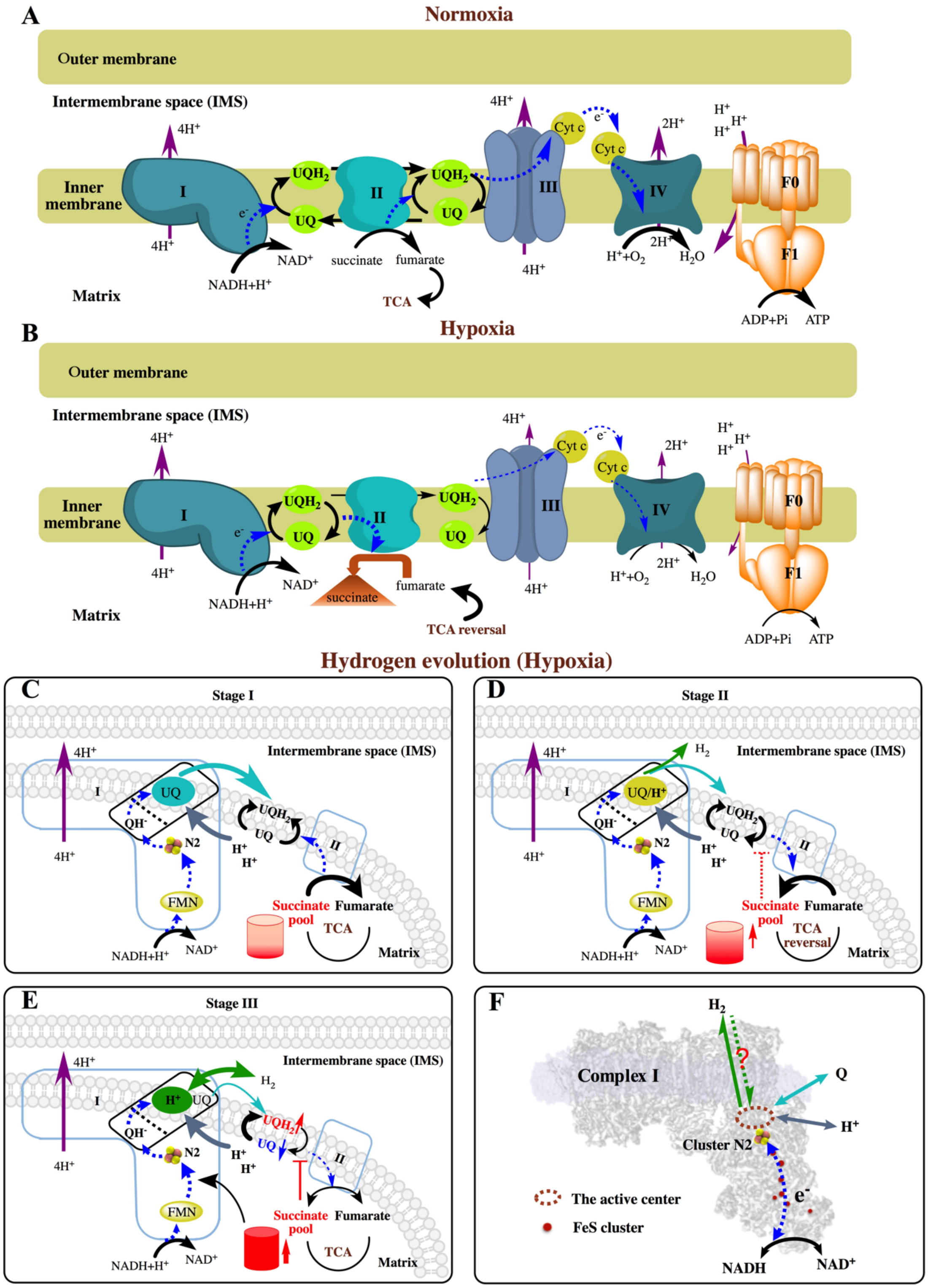

Figure 4 The schematic diagram of $H_2$ evolution mechanism in mitochondria. A-B, The schematic diagram of ETC in normoxic (A) or hypoxic conditions (B). C, Normal TCA cycle in stage I of $H_2$ evolution. D, Succinate pool begins to accumulate in stage II and start $H_2$ production. E, Succinate accumulates to a certain amount and large amount of $H_2$ produces. F, The active center of $H_2$ evolution near N2 cluster in complex I.

## Discussion

For a long time, whether mitochondrial complex I, closely related to hydrogenases structurally, do posses the $H_2$ metabolism activities attracted a lot of speculation. Here we show the experimental evidence that mitochondrial complex I is closely related to $H_2$ production to adapt to the metabolic stress such as succinate accumulation, increased reducibility of Q-pool and $NADH/NAD^+$ ratio caused by hypoxia, it is possible that exogenous $H_2$ also acts at complex I to affect redox balance, and these responses might the mechanism of why $H_2$ could greatly alleviate the oxidative damage. It is an more effective antioxidant mechanism through coupling of hydrogenase activity of mitochondrial complex I with metabolic regulation, the process like a fine-tunning oxidation-reduction valve instead of removing the reactive oxygen species after they are produced, as proposed by Dole et al(Dole et al., 1975) and Ohsawa et al(Ohsawa et al., 2007). This may be a protective mechanism of redox homeostasis regulation during long-term evolution, for example plants can adapt to continuous or intermittent rhizosphere anaerobic environments to varying degrees. Exogenous $H_2$ has positive botanical effects on plant growth and abiotic stress adaptability, e.g. high salinity(Xie et al., 2012), cadmium(Cui et al., 2013), paraquat exposure(Jin et al., 2013), heat stress(Chen et al., 2017), osmotic stress(Su et al., 2018), and drought tolerance(Jin et al., 2016). Also, the changes of metabolic intermediates conversely cause a series of adaptive regulations, such as the stabilization of HIF-1, the inhibition of α-ketoglutarate-dependent and non-heme iron-dependent dioxygenases. Succinate and fumarate have been shown to potently inhibit the expression of ten eleven translocation (TET) family of epigenetic regulator(Dando et al., 2019; Lio et al., 2019; Piccolo & Fisher, 2014).

According to the hydrogen hypothesis for the first eukaryote, the symbiont's $H_2$-metabolism function might be retained in mitochondria and readily recruitable. During the reversion to anaerobic energy metabolism, electron transport to $O_2$ was inhibited and mitochondria return to the original state and function as hydrogenase to produce $H_2$, as in some anaerobic mitochondria(Brigitte et al., 2005). We speculate that $H_2$ metabolism by complex I might be the innate function of eukaryotes during long-term evolution.

Hydrogen atom is made up of an electron and a proton. The flow of energetic electrons and protons is the essence of life. Mitochondria in eukaryotes acts as the carrier of flowing electrons and protons, the hydrogenase-like activity in mitochondrial complex I offers a huge imaginable space for biological significance.

There are still lots of questions to be resolved. For example, where is the exact catalytic center of $H_2$ production? What are the pathways for conduction of $H_2$, protons and electrons to the catalytic center? Is complex I a bidirectional hydrogenase? Is the $H_2$ metabolism of animal complex I the same as plants? Is $H_2$ metabolism an intrinsic function in eukaryotes? What is the relationship of the hypoxia activated hydrogenase activity and the numerous physiological functions reported?

## Materials and methods

**Plant Material and Growth Conditions**. Mung bean (*Vigna radiata* [L.] R. Wilczek) and soybean (Glycine max) was sterilized in 2% NaClO for 15 min, 75% alcohol for 5 min and washed three times with sterile dd$H_2O$. Then the seeds were incubated in sterile dd$H_2O$ at 55 ℃ for 20 min and soaked for 8 hours in darkness at 25±1°C. The seeds were laid flat in a seedling box covered with 2 layers of sterile gauze and incubated at 25±1°C in darkness for the required time. The roots, hypocotyls, cotyledon and true leaves of mung bean were collected at 72h for the detection of tissues hydrogen evolution. The seeds of Arabidopsis (*Arabidopsis thaliana*) *Col-0* were sterilized by 75% alcohol for 5 min and washed by sterile ddH$_2$O for three times. Arabidopsis were cultured sterilely in a Murashige & Skoog basal (MS) medium with a 16/8-h light/ dark cycle (23°C/18°C). The roots of one-week-old *G. max* and two-week-old Arabidopsis were harvested for mitochondria isolation.

**Separation of Different Cell Components**. The method of differential centrifugation was used to separate different cell components. All the operations of extraction processes were performed at 4°C. The 100 g hypocotyls of 72h *V. radiata* seedlings were collected and homogenated in precooled 300 ml extraction buffer A (25 mM Tris-Mes (pH 7.2), 5 mM EGTA, 250 mM sucrose, 1 mM MgSO$_4$, 0.5% bovine serum albumin (BSA), 0.5% poly (vinylpyrrolidone) (PVP), 10% glycerol, 1 mM phenylmethylsulfonyl fluoride (PMSF), and 1 mM dithiothreitol (DTT). PMSF and DTT were added before use). The homogenate was filtered through 4 layers of sterile gauze and then centrifuged at 1500 g for 20 min. The sediment, which contains tissue fragments, nuclei, complete cells and other heavier cell components was discard. The supernatant was centrifuged at 12000 g for 20 min. The supernatant (Plasma membrane, microsomes, endoplasmic reticulum, cytosol, *et al*) and pellets (mitochondria, lysosome, peroxisome, *et al*) were collected respectively. The pellets suspended by 1mL suspension buffer B (5mM potassium phosphate buffer (PH 7.2), 5 mM KCl, 250 mM sucrose, 0.5% (W/V) BSA and 1mM DTT which was added before using). Part of 12000 g supernatant was centrifuged at 80000 g for 30 min using Beckman ultra-centrifuge (Beckman Coulter, Inc., USA). Supernatant and pellets (including plasma membrane,) were collected respectively and precipitate was suspended in 1 mL suspension buffer B. Protein concentration was measured using Bradford protein assay kit (Tiangen Biotech, Beijing Co., LTD, CN).

**Preparation of Crude Mitochondria**. The crude mitochondria were prepared as described previously with some modifications(Douce et al., 1987). Hypocotyl (300 g) of 72h seedlings were homogenized in 900 mL of extraction buffer containing 0.3 M D-mannitol, 4 mM L-cysteine, 1 mM EDTA-Na$_2$, 0.1% (w/v) BSA, 0.6% (w/v) PVP and 30 mM 3-(N-morpholino) propanesulfonic acid (MOPS), pH to 7.5 with NaOH). Hypocotyls were homogenized at low speed for three 5-seconds by Warning blender (Waring laboratory). After filtered through 4 layers of gauze, the filtrate was adjusted to pH 7.5 and centrifuged at 1500 g for 20 min. The precipitation contains nucleus, intact cells and other heavier cellular components was discarded. The supernatant was centrifuged at 12000 g for 20 min, and pellet was suspended with 4 mL washing medium containing 0.3 M D-mannitol, 0.1% (w/v) BSA, 1mM EDTA-Na$_2$, and 10 mM MOPS (adjusted pH to 7.2 with NaOH). All the resuspended solution was collected, centrifuge at 1500 g for 10 min again and pellet was discarded; The supernatant was centrifuged at 12000 g for 20 min. The pellet containing crude mitochondria was gently suspended in a small volume of washing medium (according to the experiments to decide whether added with 1mM EDTA-Na$_2$ or not). Then crude mitochondria were sub-packed and quick-frozen with liquid nitrogen and stored at - 80 C. All the above operations of extraction processes were performed at 4°C. Protein concentration was measured using Bradford protein assay kit (Tiangen Biotech, Beijing Co., LTD, CN). The mitochondria integrity was assessed by O$_2$ electrode of Hansha oxytherm (Hansatech Instruments) and the criteria was according to respiratory control rate (RCR, it is the respiratory rate in the presence of added ADP compared to the rate obtained following its expenditure)(Jacoby et al., 2015).

**Measurement of $H_2$ Evolution.** $H_2$ evolution was monitored in a total volume of 500 μL in 1.9 mL gas chromatographic flasks containing 200-500 μg cell components or crude mitochondria and $H_2$ evolution buffer (HEB) (0.3 M mannitol; 0.1% (w/v) BSA; 10 mM MOPS, pH 6.0). The HEB without cell components or mitochondria were regarded as the negative control. For the $H_2$ evolution detection of seedlings or tissues, the seedlings or tissues randomly selected were added into 40 mL (seedlings) or 22.5 mL (tissues) gas chromatographic flasks containing 2 mL sterile dd$H_2O$ and sealed quickly. There were 10 samples each bottle and 5 repetitions each time point. The headspace of GC flask was flushed with $N_2$ for 2 minutes to remove $O_2$. Then the mitochondria samples were placed in an artificial climate incubator at 25℃ or 35℃ in darkness for needed time. The cell components, seedlings or tissues were incubated at 25℃ for 16h in darkness.

For analyzing endogenous $H_2$ content, headspace gas chromatography (GC) (Shimadzu, GCMS-QP2010S) was adopted. The chromatographic system was composed of a GC equipped with a barrier discharge ionization detector (BID) and a column containing the SH-Rt-Msieve 5A stationary phase (30 m, 0.32 mm internal diameter, 30 μm film; Shimadzu). The sample loading amount was 1 mL. The analytical conditions were as follows: helium as carrier gas, the constant linear velocity of 60 cm/s in column, the column temperature of 50℃, injection port temperature of 250℃, split ratio of 5:1, pressure with 200.2 kPa and the column flow rate with 3.85 mL/min. Then $H_2$ evolution activities were calculated as nmol/h/mg protein or nmol/h/g fresh weight.

To study the effects of different buffers on $H_2$ evolution, the 500 μg crude mitochondria was suspended in suspension buffer B or washing medium. Then they were placed into extraction buffer A, suspension buffer B or washing medium respectively for 14h in darkness at 25℃. To study the effects of EDTA on mitochondrial $H_2$ production, we added mitochondria into the HEB with or without EDTA-$Na_2$. To identify the favorite reaction temperature of mitochondria $H_2$ evolution, samples were placed at 25 ℃, 30 ℃, 35 ℃, 40 ℃ or 45 ℃ in darkness for 4 hours. To test the optimal pH, the mitochondria samples were suspended into HEB（adjusted to pH 4.0, 5.0, 6.0, 7.0, respectively）and incubated at 35 ℃ in darkness for 4 hours. Before GC detection, mitochondria samples incubated for 3h, 4h or 5h were treated by shocking at 400 rpm at 40℃ for 10 min to release all the $H_2$ from the liquid. No shocking treatment was as control to detect the influence of shocking for $H_2$ detection. Two or three repeats for each treatment were carried out. All the mitochondrial samples were tested under hypoxic condition and $H_2$ abundance were detected by GC method mentioned above.

**Analysis of Mitochondrial Metabolites**. The mitochondria (500 μg) were placed in a 1.9 mL gas chromatographic flask. HEB supplemented the reaction volume to 500 μL. The headspace of GC flask was flushed with $N_2$ for 2 minutes to remove $O_2$. Followed by incubating the samples for 0 h, 6 h and 10 h at 25℃ under normoxic or hypoxic condition or 0-7h at 35℃ under hypoxia. Then immediately freezed with liquid nitrogen and stored at -80 ℃. The samples treated at 25℃ were sent to Shanghai Applied Protein Technology Co., Ltd. (dry ice transportation) for analysis of targeted energy metabolites using high performance liquid chromatography-mass spectrum (HPLC-MS). The samples treated at 35℃ were analyzed by MS in Beijing university of technology. The Detailed methods were as follows:

**Extraction of Metabolites**. After thawing on ice, the samples were mixed by vortexing. Each sample of 200 μL was added into 800 μL precooled methanol (Merck) / acetonitrile (Merck) (1:1, v/v) and mixed by vortexing. The mixture was ultrasounded for 20 min in ice bath, incubated at -20 ℃ for 1 h to precipitate protein, centrifuged at 14000 g for 20 min at 4℃. The supernatant was taken and vacuum dried. Then the dry powder was redissolved into 100 μL acetonitrile-water solution (1:1, v/v) and centrifuged at 14000 g for 20 min at 4℃. The supernatant was analyzed for next detection.

**HPLC-MS Analysis of Mitochondria Treated at 25℃**. The samples were separated by liquid chromatography (LC) using Agilent 1290 Infinity LC HPLC (Agilent). ACQUITY UPLC BEH amide column (1.7μm, 2.1 mm×100 mm) from Waters was used. Mobile phase. A: 10 mM ammonium acetate aqueous solution. B: acetonitrile. Samples were placed in the automatic sampler at 4 ℃. Injection volume: 6 μL. Column temperature: 45 ℃. Flow rate: 300 μL /min. Gradient of the related liquid phase: 0-18 min 90% to 40% B, 18-18.1 min, 40% to 90% B, 18.1-23 min, 90% B. The mixture of standard metabolites were set up in the sample queue for correction of chromatographic retention time. A 5500 QTRAP MS (AB SCIEX) was used for MS analysis under negative ion mode. 5500 QTRAP ESI source conditions were as follows: source temperature: 450 ℃, ion Source Gas1(Gas1): 45, ion Source Gas2 (Gas2): 45, Curtain gas (CUR): 30, ionSapary Voltage Floating (ISVF): -4500 V.

**MS Analysis at 35℃**. Thermo Orbitrap UHPLC-MS (Thermo) was used to determine the changes of metabolic substrates at 35℃. The UHPLC parameters were as follows: mobile phase: acetonitrile and water (80： 20, v/v). velocity of flow: 200 μL/min. The injection volume: 10 μL. All MS detection was performed on a Thermo LTQ Orbitrap XL MS under negative ion modes. The optimal MS parameters were set as follows: capillary temperature: 320 °C. Resolution: 30000, AGC target: 2*10^5, maximum inject time: 500 ms. LTQ Orbitrap XL MS was calibrated using commercially available calibration solutions prior to measurements. Qualitative analysis was obtained by secondary MS using collision-induced dissociation (CID) method and compared to the ion spectrum of thermo database. For further qualitative analysis of metabolic substrate, second spectrometry was used. The parents (m/z 145.01) analysis for α-ketoglutarate was performed with 22 eV collision energy. The parents (m/z 117.01) analysis for succinate was performed with 30 eV collision energy. The parents (m/z 115.00) analysis for fumarate was performed with 25 eV collision energy. The parents (m/z 133.01) analysis for malate was performed with 26 eV collision energy.

**Data processing**. Multiquant software was used to extract chromatographic peak area and retention time in Shanghai Applied Protein Technology co.ltd. The software of Thermo Xcalibur Qual Browser 2.1.0 was used to extract chromatographic peak area and retention time of the samples treated by 35℃. The standard metabolites were used to correct retention time and identify metabolites. The mzCloud Dataviewer (www.mzcloud.org) was used to search characteristic product ions in the secondary MS.

**Effects of $O_2$ on Mitochondrial $H_2$ Evolution**. The $H_2$ evolution reaction system of mitochondria was the same as above. The $O_2$ was removed by $N_2$ and not excluded group was as control. Then the samples were cultured in dark for 2 h, 4 h, 6 h, 8 h, 10 h, 12 h at 25℃ or 1 h, 2 h, 3 h, 4 h, 5 h, 6 h, 7 h, 8 h at 35 ℃. $H_2$ content was determined by headspace GC as mentioned above, each treatment was repeated three times. The assays were repeats for at least twice.

**Effect of Metabolic Intermediates on Mitochondrial $H_2$ Evolution**. Malate and pyruvate can provide endogenous NADH for complex I. Succinate and fumarate are metabolic substrates and products of complex II. Rotenone, inhibitor of complex I, blocks the transfer of electrons from NADH to UQ. Malonic acid is a succinate analogue, which can competitively inhibit complex II activity. In order to explore the relationship between complex I and $H_2$ metabolism, we added the substrates or inhibitors to the $H_2$ production system. All the groups were as follows: 500 µg mitochondria (control), 500 µg mitochondria +5 mM malate + 5 mM pyruvate and 500 µg mitochondria +5 mM malate + 5 mM pyruvate +5 μM rotenone for studying the relationship of complex I and $H_2$ evolution. 500 µg mitochondria +10 mM succinate; 500µg mitochondria +5 mM fumarate, 500 µg mitochondria +10 mM succinate + 5 mM fumarate and 500 µg mitochondria +10 mM succinate+ 5 mM fumarate +10 mM malonic acid for researching the relationship of complex

II and $H_2$ evolution, adjusted the pH of all the reaction systems to 6.0 and the volume to 500 μL with HEB. Hypoxia was achieved as the same method as mentioned above. Samples without inhibitors and substrates were used as control. After 3h incubation at 35℃, $H_2$ content was determined by headspace GC as mentioned above (n=3 for each group and the assays were repeated at least for twice).

**Effect of Mitochondrial Electron Transfer Inhibitors on $H_2$ Evolution**. Rotenone, and malonic acid are inhibitors of complex I or II respectively as mentioned above. Antimycin A is an inhibitor of complex III, which blocks the electron transfer from $UQH_2$ to cytochrome c. $NaN_3$ inhibits the electron transfer from cytochrome b to cytochrome $c_1$ in complex IV. Oligomycin can block proton channel of F0 of F0/F1 ATPase, inhibit proton reflux into matrix through the inner membrane and lead to no ATP synthesis. FCCP is uncoupling agent of oxidative phosphorylation, which destroys the proton gradient of mitochondrial inner membrane, leads to uncouple the electron transfer and ATP synthesis. Rotenone, antimycin A, oligomycin and FCCP were dissolved in DMSO, whereas malonic acid and $NaN_3$ were dissolved in HEB. The reaction system (500 μL) included 300-500 μg mitochondria, 5 μM rotenone, 10 mM malonic acid, 10 μM antimycin A, 5 μM $NaN_3$, 2 μM oligomycin or 25 μM FCCP, adjusted the volume to 500 μL with HEB. Hypoxia was achieved as above. Samples without inhibitors were used as control. After 3h incubation at 35℃, $H_2$ content was determined by headspace GC mentioned above (n=3 for each group and the assays were repeated for thrice).

**The Change of pH Values During $H_2$ Evolution.** The $H_2$ evolution system of mitochondria were incubated at 35℃ for 1h, 2h or 3h. The pH values of each sample were detected by pH electrode of Hansha oxytherm (Hansatech Instruments). (n=3 for each group and the assays were repeated for twice).

**Effect of UQ on Mitochondrial $H_2$ Production**. In order to further explore the relationship between $H_2$ production activity and complex I II, Coenzyme $Q_{10}$ (Sigma) was added into the reaction system. The groups were as follows: 500 μg mitochondria as the control; 500 μg mitochondria +5 mM malate +5 mM pyruvate +10 mM succinate+5 mM fumarate; 500 μg mitochondria +5 mM malate +5 mM pyruvate +10 mM succinate +5mM fumarate+60 μM UQ; 500 μg mitochondria +5 mM malate +5 mM pyruvate +10 mM succinate +5mM fumarate+ UQ +5 μM rotenone; 500 μg mitochondria +5 mM malate +5 mM pyruvate +10 mM succinate +5mM fumarate+ UQ +10 mM malonic acid, adjusted all the systems pH to 6.0. The reaction volume was supplied to 500 μL with HEB. $H_2$ content was detected by GC method as the same as above.

**NADH Content Changes During $H_2$ Evolution.** In order to determine the changes of NADH content during $H_2$ production, colorimetric method was used. The NADH content was measured according to the manual of AAT Bioquest's Amplite™ Colorimetric NADH Assay Kit. The NADH probe is a chromogenic sensor that has its maximum absorbance at 460 nm upon NADH reduction. The NADH probe can recognize NADH and the signal can be easily read by an absorbance microplate reader at 460 nm. After mitochondria was mixed in $H_2$ production systems for 0-7h, $H_2$ evolution mixture of 50 μL were added into 50μL working solution, incubated at room temperature for 1 hour and protected from light. Then monitored the absorbance increase with an absorbance plate reader at 460 nm. The stand curve was obtained by a series of NADH standard solution.

**Statistical Analysis**. GraphPad Prism 6.0 was used for data analysis and figures creation. ChemDraw professional was used to draw the the mechanism diagram of $H_2$ evolution. The measurements were taken from distinct samples. P values were calculated using unpaired t-test (two-tailed) for pairwise comparisons or ordinary one-way ANOVA test for multiple comparisons followed by Dunnett's multiple comparisons test, unless otherwise stated. Data were expressed as mean±SD. Differences with P-values<0.05, <0.01, < 0.001 or < 0.0001 were considered significant.

## Compliance and ethics

The authors declare that they have no conflicts of interest with the contents of this article.

## Acknowledgments

Thanks to all the members of our research team for their help during the research. This work was supported by Special Fund for Agro-scientific Research in the Public Interest (201303023), Military Logistics Key Open Research Projects (BHJ17L018) and Chaoyang District Postdoctoral Research Foundation (2018ZZ-01-11).

**Supplementary Materials:**

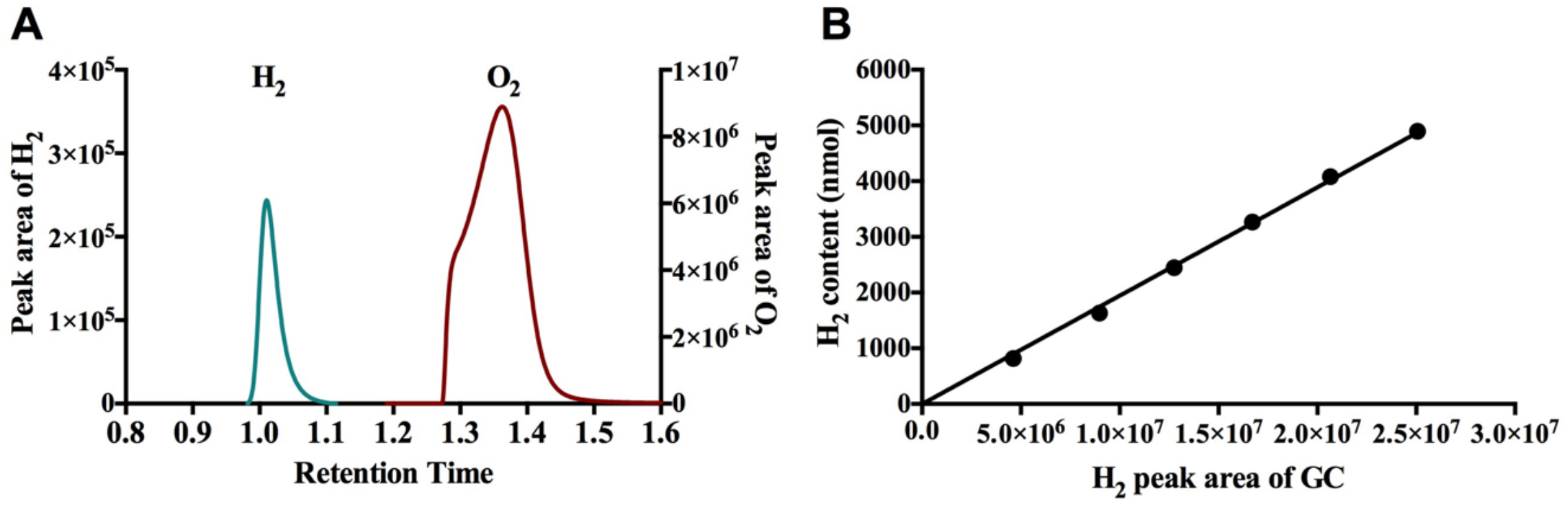

Figure S1. $H_2$ and $O_2$ were detected by GC. A, $H_2$ and $O_2$ could be clearly separated by GC. B, The linear relationship between $H_2$ content and GC peak area ($R^2$=0.9986).

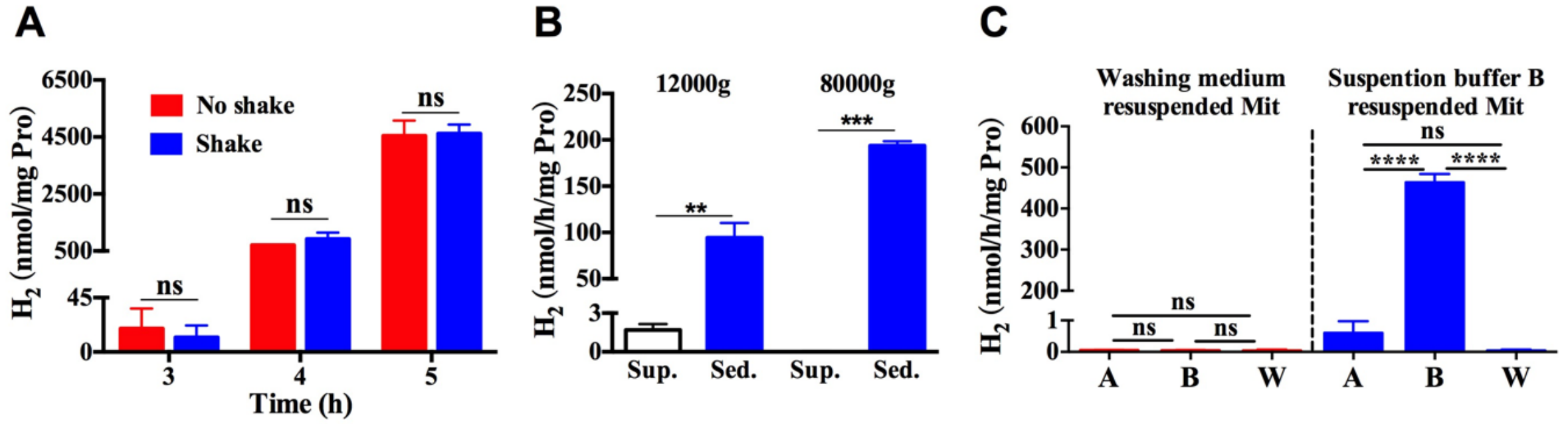

Figure S2. Detection conditions of $H_2$ production. A, Effect of pre-test oscillating treatment on $H_2$ detection in mitochondria of *V. radiata* hypocotyls (mean±SD, n=2). B, $H_2$ production by different cell components of *V. radiata* hypocotyls separated by differential centrifugation (mean±SD, n=2). C, Effect of different buffers on $H_2$ evolution activity of mitochondria (mean±SD, n=3). A: extraction buffer A (25 mM Tris-Mes, 5 mM EGTA, 250 mM sucrose, 1 mM $MgSO_4$, 0.5%(W/V) BSA, 0.5% PVP, 10% glycerol, 1 mM PMSF, and 1 mM DTT, pH 7.2), B: suspension buffer B (5mM potassium phosphate buffer, 5 mM KCl, 250 mM sucrose, 0.5% (W/V) BSA and 1mM DTT, pH 7.2), W: washing medium (0.3 M D-mannitol, 0.1% (w/v) BSA and 10 mM MOPS, 1mM EDTA, pH 7.2). ns, no significant. **, $P<0.01$, ***, $P<0.001$, ****, $P<0.0001$.

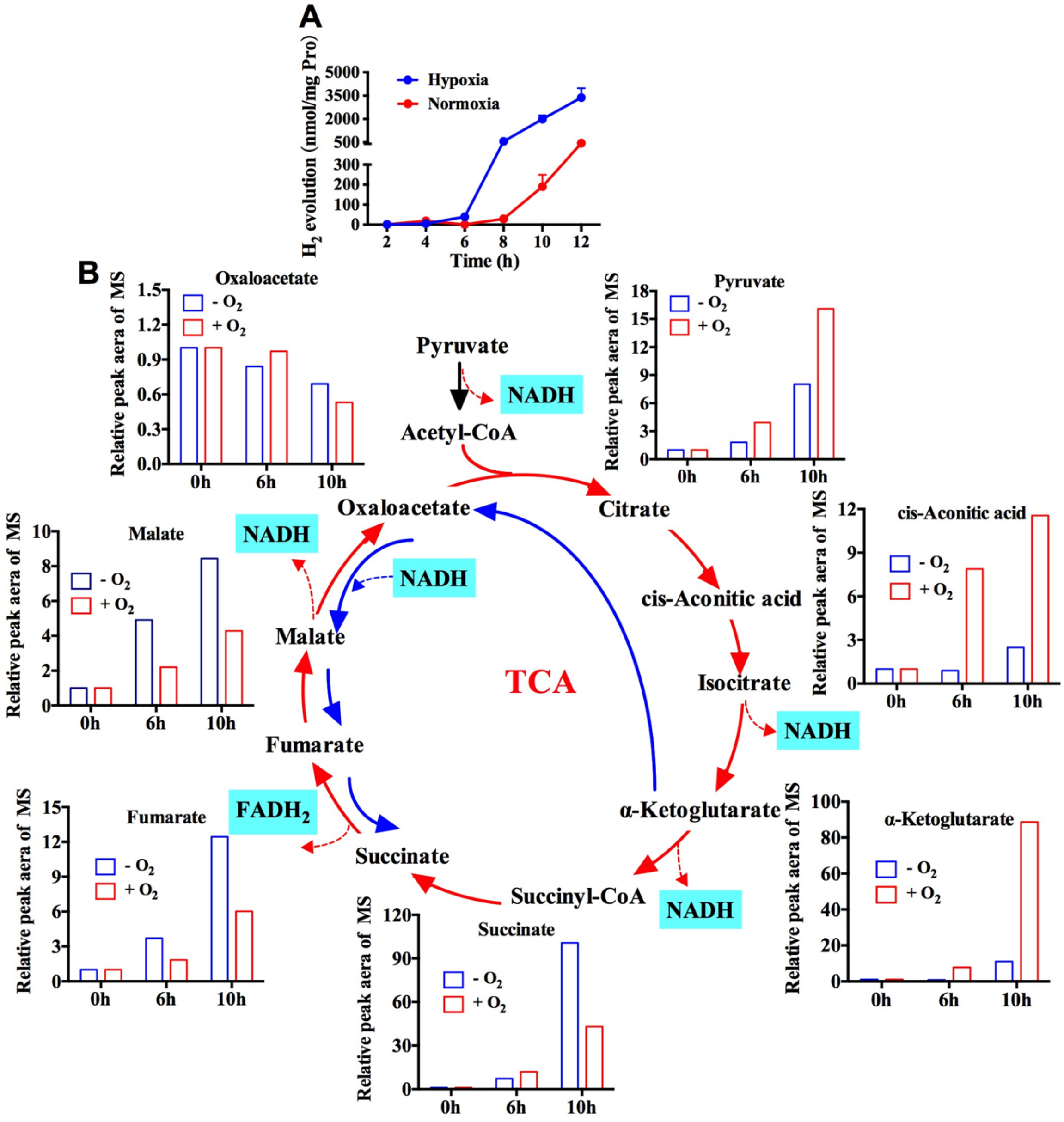

Figure S3. Metabolome analysis of *V. radiata* hypocotyls mitochondria during $H_2$ production at 25℃. A, Time course of mitochondria $H_2$ evolution (mean ± SD, n=3). B, Metabolome analysis of mitochondria under normoxic and hypoxic conditions.

**A α-Ketoglutarate**

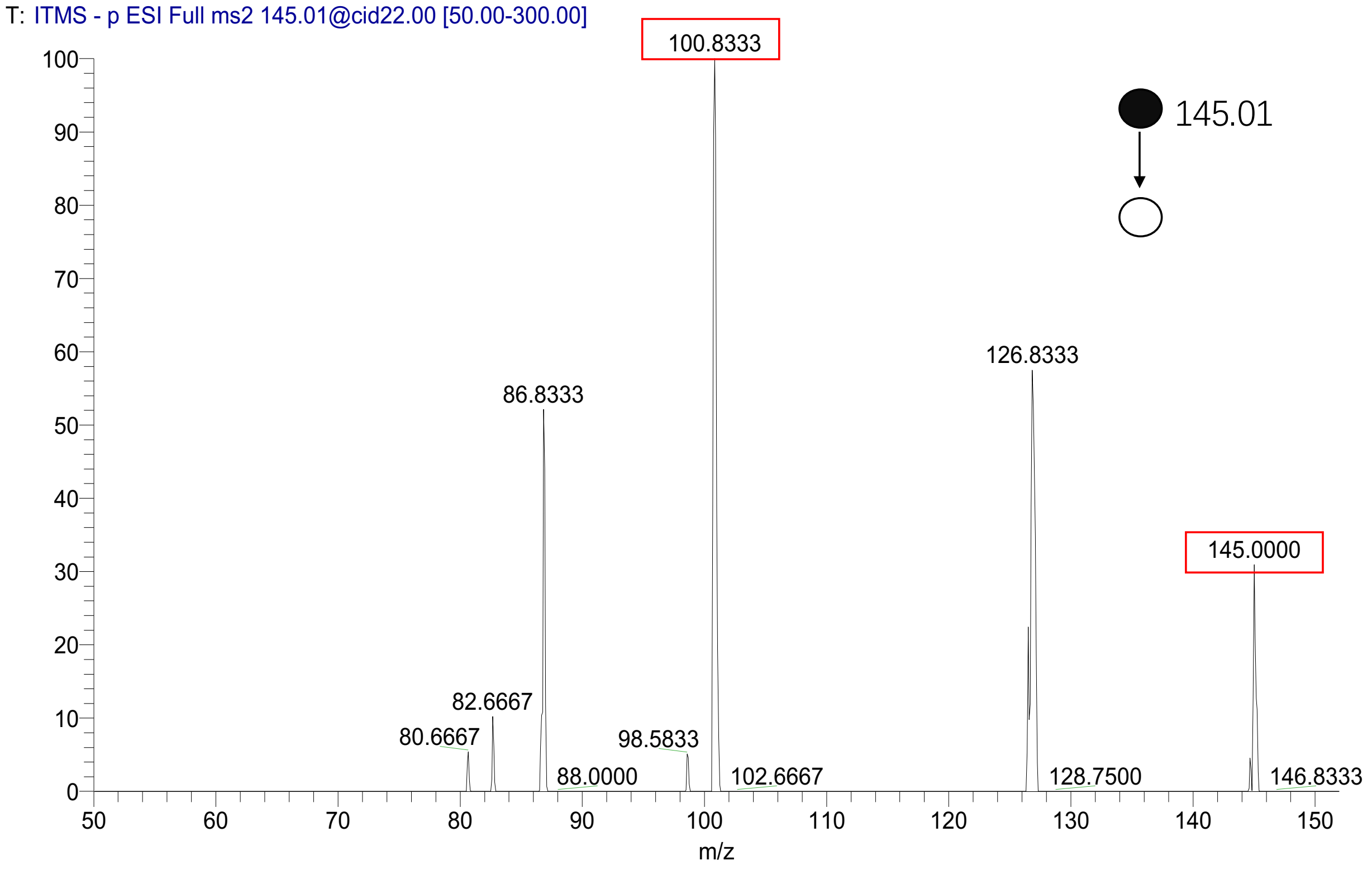

**B Succinate**

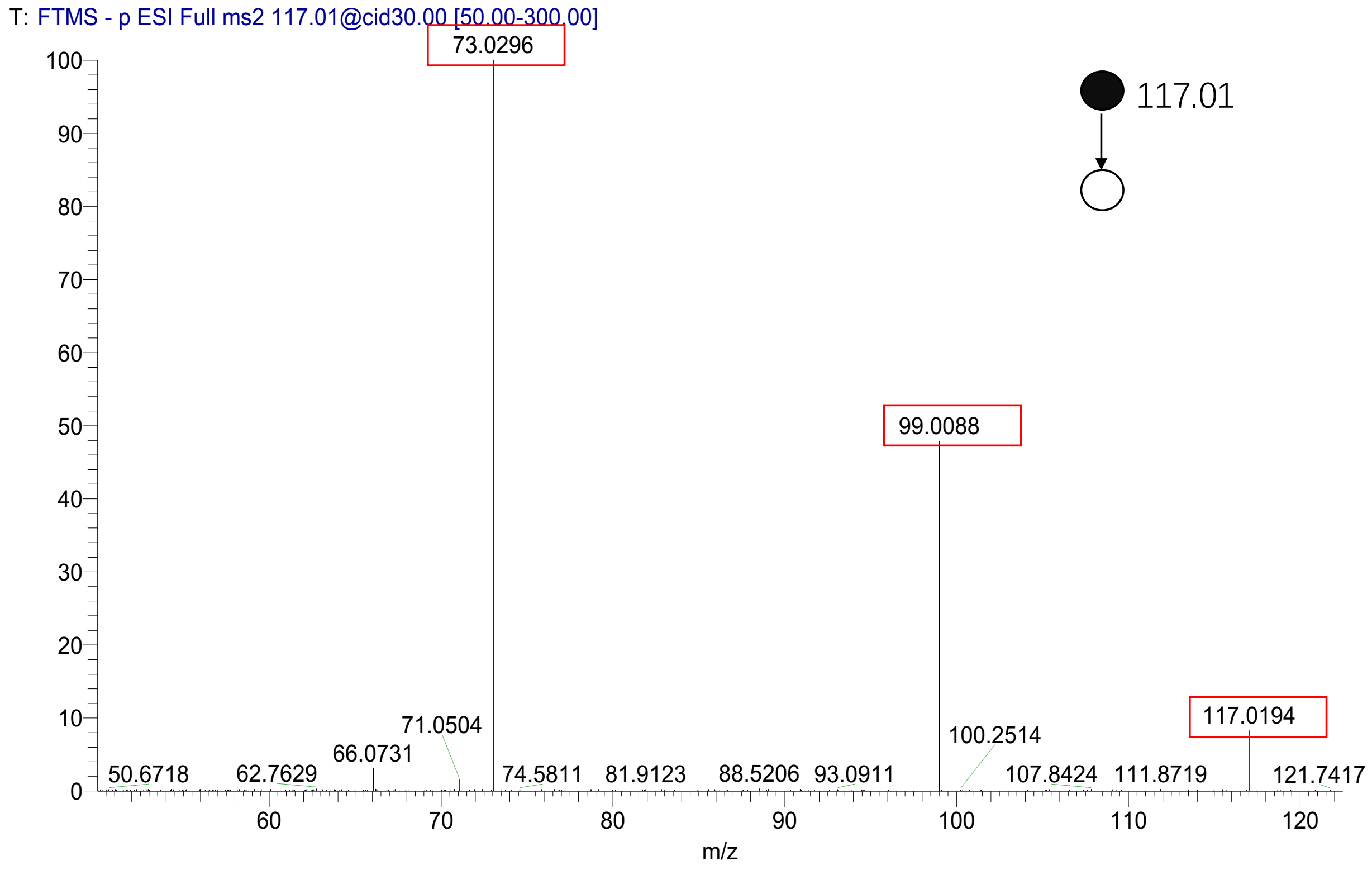

**C Fumarate**

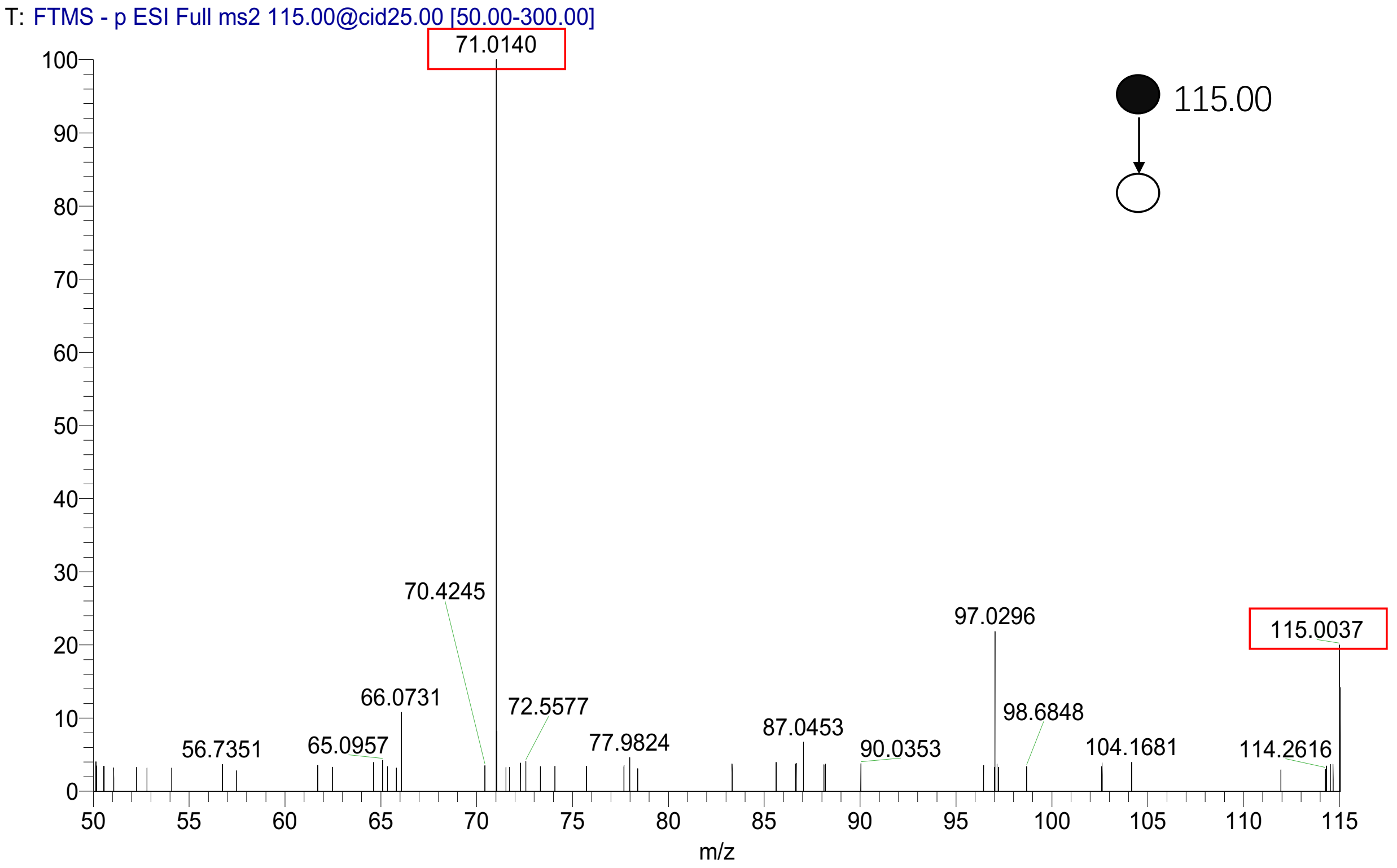

**D Malate**

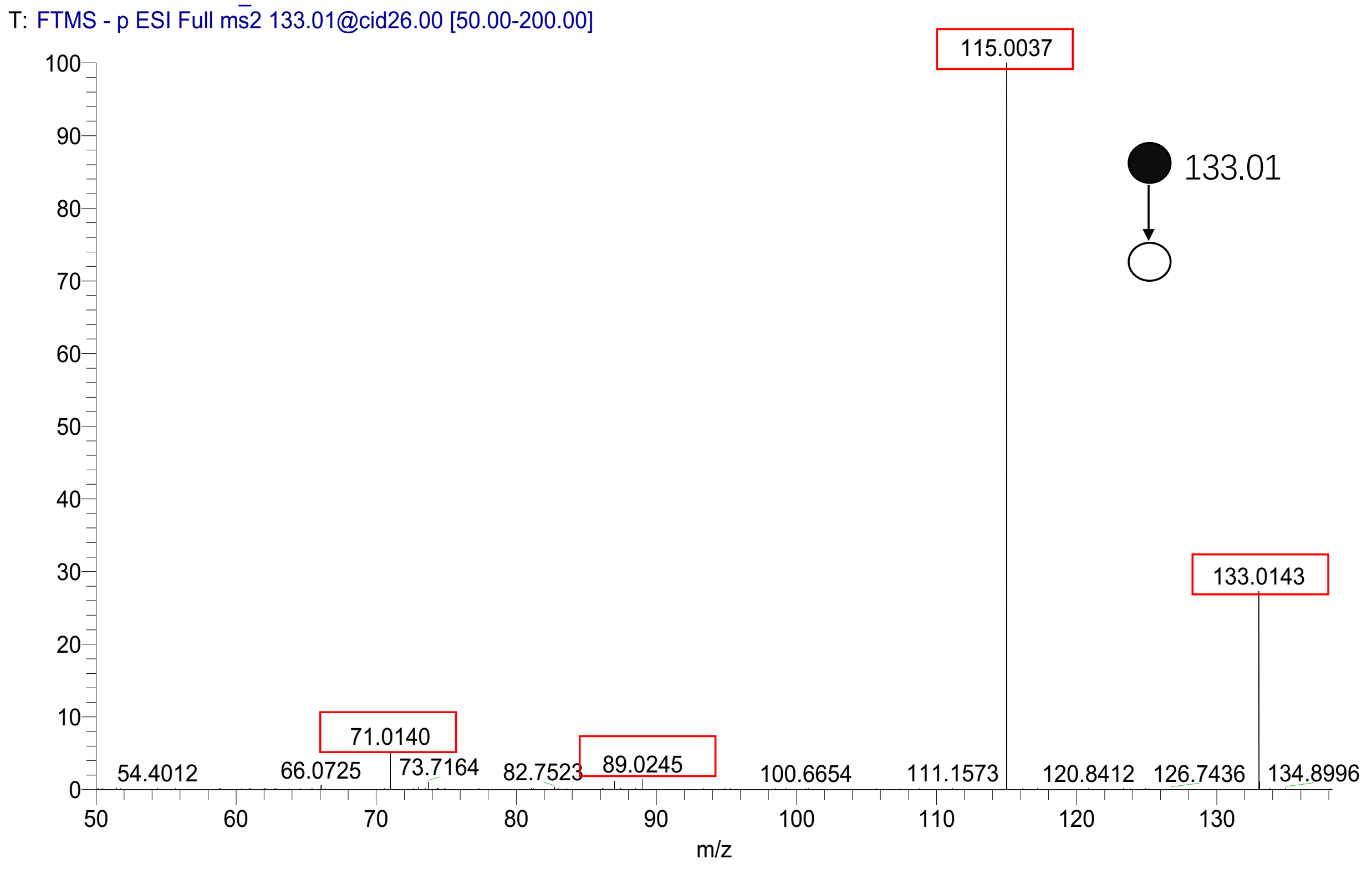

Figure S4. The secondary MS of mitochondrial metabolites. A-D, The secondary MS of α-ketoglutarate (A), succinate (B), fumarate (C) and malate (D).

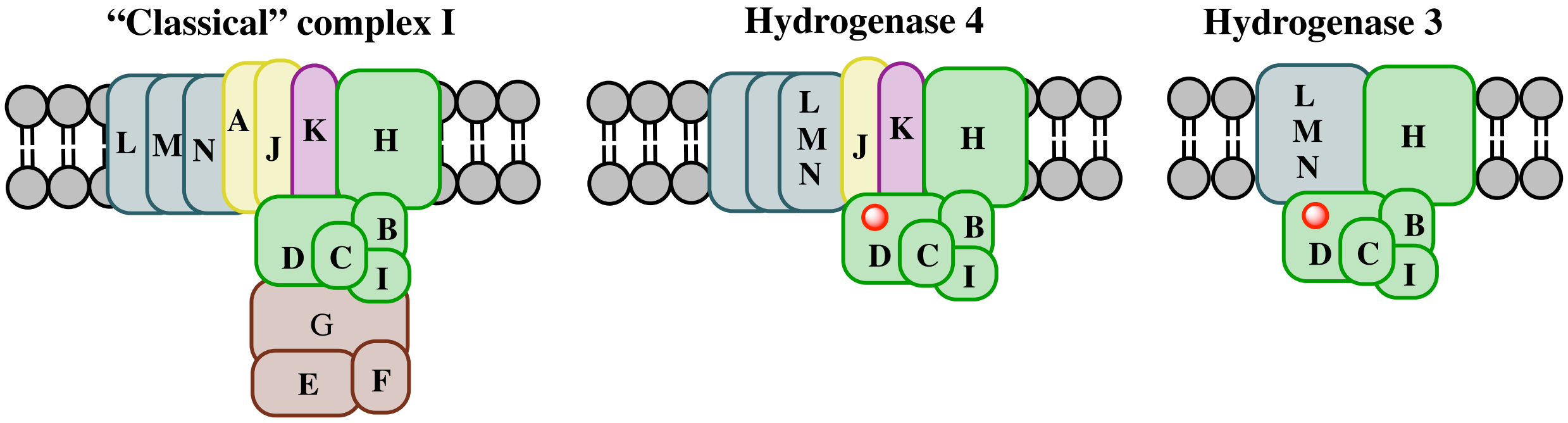

Figure S5. Schematic homology representation of the complex I and membrane bound hydrogenases 4 and hydrogenases 3(Moparthi & Hagerhall, 2011). The quinone module (Q-module) of complex I is composed of NuoC, NuoI, NuoB and NuoD. Two proteins resembling the small and the large subunit of NiFe-hydrogenases, which in complex I correspond to the NuoB and NuoD subunits, respectively. The Q-module accepts electrons from the N-module (NADH dehydrogenase module, compose of NuoE, NuoF and NuoG) and transfers them via iron–sulfur clusters to UQ. The NiFe hydrogenases active site is indicated as a red circle.